\documentstyle[12pt]{article}
\begin{document}
\begin{center}
{\Large THE EXACT DETERMINATION OF THE\\
COSMOLOGICAL CONSTANT}
\vskip 8mm
{\bf V.V.Burdyuzha}\\
{\it Astro Space Centre Lebedev Physical Institute of Russian }\\
Academy of Sciences\\
Profsoyuznaya 84/32, 117810 Moscow, Russia

\end{center}

\vskip 12mm

\begin{abstract}
With cooling of the cosmological plasma during phase transitions in the
early Universe vacuum condensates of quantum fields were produced with
negative energy density. Probably these condensates have compensated the
initial vacuum energy. The present vacuum is the vacuum condensate of the
last relativistic phase transition ($T_{cr}\sim 100$ MeV) (its information
is carried by pseudo-Goldstone bosons ($\pi $-mesons)). This allows us to
use Zeldovich's formula that can calculate exactly the value of the
cosmological constant in the present epoch. If the most likely values  
$\Omega_\Lambda $ are 0.7-0.8 then $H_0$ falls in the range from 72.5 to 
67.8 (km/s)/Mpc.  
\end{abstract}

$\cdot $\vspace*{1cm}

The cosmological constant problem is one of the intriguing problems of
modern physics and cosmology. The suggestions for its solution ${}^{1-10}$
have attracted a lot of attention (for a review see article ${}^2$). In this
notice I provide some arguments and calculate the present-day value of
cosmological constant as the vacuum energy density of the last relativistic
phase transition.

The terms of vacuum energy and cosmological constant ($\Lambda$-term) are
used practically synonymously in modern cosmology.

The physical vacuum is a complex heterogeneous system of classical and
quantum fields. It has the interior structure and changes its state with the
change of existence conditions (a phase transition for example). The reality
of vacuum energy forces to insert it in equations of the gravitational
theory in the form of $\Lambda $-term. $\Lambda $-term was introduced first
by A.Einstein in his equations in order to obtain static cosmological
solutions. It was also named the cosmological constant

$$
R_{\mu \nu }-\frac 12 g_{\mu \nu }R+\Lambda g_{\mu \nu } =8\pi GT_{\mu \nu }%
\eqno(1)
$$

Here $R_{\mu \nu },R,g_{\mu \nu }$ and $T_{\mu \nu }$ are Riemann, Richi,
metric and energy momentum tensors. $G$ is the Newtonian gravitational
constant. Up till now these equations were the central ones in cosmology.
The equation (1) leads to the Friedmann equations from which the density
parameter of the Universe $\Omega _o$ can be defined.

Currently some cosmological data point out that we live in the Universe with
a nonzero vacuum energy density ${}^{7}$ (the lensing optical depth of
galaxies at the moderate redshifts is substantially greater than in
Einstein-de Sitter Universe ${}^{11}$). The "age crisis", the problem of
large-scale structure formation, the spatial curvature are solved better if $%
\Omega_{\Lambda} = \rho_{\Lambda}/ \rho_{cr} \equiv \frac{\Lambda c^{2}}{3
H^{2}_{o}} \ne 0 \;$ (here $\Omega_{\Lambda}$ is the vacuum energy density
in units of the critical density $\rho_{cr} = \frac{3 H^{2}_{o}}{8 \pi G},
\; H_{o}$ - is the Hubble constant). $\Omega_{\Lambda} = \Omega_{o} -
\Omega_{m}$ is one way to resolve the discrepancy between $\Omega_{o}
\approx 1$ ($\Omega_{o}$ should be the order of unity since the spatial
curvature is practically zero in the present epoch) and the density of
matter $\Omega_{m} = \Omega_{b} + \Omega_{CDM} + \Omega_{HDM} = 0.2-0.3 \;
{}^{12}$. Here $\Omega_{b}, \Omega_{CDM}, \Omega_{HDM}$ are the densities of
baryons, cold and hot dark matter particles respectively (also all are in
units of the critical density).

To be exact it is pertinent to say here few words about the birth of the
Universe and its early evolution. It is reasonable to suppose that our
Universe nucleates spontaneously out of "nothing" ${}^{13-16}$ either in a
clean vacuum state ${}^{8,17}$ or in an anisotropic state with some number
of particles and some nonequilibrium state of vacuum ${}^{18}$. Probably the
processes of relaxations took place near Planck parameters. The creation of
the three-dimensional Universe (as one of these processes) may be the
process of spontaneous breaking of the local supersymmetry of a
high-dimensional space-time. Naturally subsequent evolution of the Universe
was accompanied by the decrease of it symmetry via relativistic phase
transitions (RPT). Possible series of RPT were:

\begin{center}
{\ 
\begin{tabular}{ccccc}
P & $\Longrightarrow$ & $D_{4} \times [SU(5)]_{SUSY}$ & $\Longrightarrow$ & $%
D_{4} \times [U(1) \times SU(2) \times SU(3)]_{SUSY}$ \\ 
& $10^{19} Gev$ &  & $10^{16} Gev$ &  \\ 
&  &  &  &
\end{tabular}
} 
\vspace*{.5cm} \\{\
\begin{tabular}{cccc}
$\Longrightarrow$ & $D_{4} \times U(1) \times SU(2) \times SU(3)$ & $%
\Longrightarrow$ & $D_{4} \times U(1) \times SU(3)$ \\
$10^{5}-10^{10} Gev$ &  & $10^{2} Gev$ &  \\
&  &  &
\end{tabular}
}
\end{center}

\vspace*{.5cm} {\
\begin{tabular}{ccc}
\hspace*{.3cm} & $\Longrightarrow$ & $D_{4}\times U(1)$ \\
& $10^{2} Mev$ &  \\
&  &
\end{tabular}
} \\\\or more complex initial stage $P \Longrightarrow E_{6} \Longrightarrow
0(10) \Longrightarrow SU(5) \Longrightarrow ... $. $P$ here is the group of
local supersymmetry joining all physical fields and interactions; $D_{4}$ is
the group of diffeomorphisms corresponding to the gravitational interaction;
$[SU(5)]_{SUSY}$ is the group of Grand Unification with the global
supersymmetry; $U(1) \times SU(2) \times SU(3)$ is the group of Standard
model symmetry of elementary particles physics.

The spontaneous compactification can be considered as the first RPT as a
result of which the gravitational vacuum condensate ${}^{18}$ was produced ($%
E \sim 10^{19}$ GeV). But this transition is practically impossible to
describe well now. Probably after the first RPT $\Lambda$-term must be in
the form: $\Lambda = \Lambda_{G} + \Lambda_{QF}$, where $\Lambda_{G}$ is the
gravitational vacuum condensate; $\Lambda_{QF}$ is the vacuum of quantum
fields (zero vibrations of quantum fields, nonperturbative condensates,
Higgs condensates). The calculation of $\Lambda_{QF}$ belongs to group of
tasks solved in the frame of renormalizable models of quantum field theory.
The rest of RPT are described by modern theories of elementary particles $%
{}^{19-20}$ (the reality of two last RPT is evident to everybody).

With the cooling of the cosmological plasma during RPT vacuum condensates
with a negative energy density are produced. These condensates have the
asymptotic equation of state $\rho _{vac}=-\epsilon _{vac}=const$. Thus RPT
series were accompanied by the generation of negative contributions in
initial $\Lambda $-term. Therefore $\Lambda $-term was changed during the
Universe evolution and can be calculated exactly since the present vacuum is
the vacuum condensate of the last RPT ($T_{cr}\sim 100$ MeV). The
condensates in the modern quantum field theory are macroscopic mediums with
quasiclassical properties. The periodic collective motions in this medium
are perceived as pseudo-Goldstone bosons. But $\pi $-mesons that were
produced as a result of this RPT are pseudo-Goldstone bosons that carry the
information about this vacuum.

Ya. Zeldovich ${}^{21}$ attempts to account for a nonzero vacuum energy
density of the Universe in terms of quantum fluctuations (the gravitational
force between particles in the vacuum fluctuations as a higher-order
effect). Thus using Zeldovich's formula ${}^{21}$ we can get the value of
the cosmological constant ($\Lambda $-term) and the vacuum energy density:
$$
\Lambda =8\pi G^2m_\pi ^6h^{-4}\simeq 1.289\times 10^{-56}\;cm^{-2}
$$
$$
\rho _\Lambda =Gm_\pi ^6c^2h^{-4}\simeq 6.908\times 10^{-30}\;gcm^{-3}%
\eqno(2)
$$

in which $m_\pi $ is the mean mass of $\pi $-mesons ($m_\pi =\frac{2m_{\pi
^{\pm }}+m_{\pi ^o}}3=138.0387\;Mev$); $h$ is the Planck constant (to use
Zeldovich's formula the author ${}^{17}$ have suggested the first). Then $%
\Omega _\Lambda =\rho _\Lambda /\rho _{cr}$ can be calculated for different
values of the Hubble constant, which is well unknown now,: \vspace*{.5cm}

\begin{center}
\begin{tabular}{|l|c|c|c|c|c|c|c|c|}
\hline
$H_{o}(km s^{-1}/Mps)$ & 50 & 55 & 60 & 65 & 70 & 75 & 80 & 85 \\ 
&  &  &  &  &  &  &  &  \\ 
$\Omega_{\Lambda}$ & 1.47 & 1.21 & 1.02 & 0.87 & 0.75 & 0.65 & 0.57 & 0.51
\\ \hline
\end{tabular}
\end{center}

\vspace*{.5cm} If the most likely values $\Omega _\Lambda $ are 0.7-0.8 then 
$H_0$ falls in the range from 72.5 to 67.8 (km/s)/Mpc. Substituting the
Planck mass into formula (2), the difference in 120 orders between the
current value and its value at the Planck time can be found. Here it is also
worth noting two important facts. Practically the total compensation of
initial $\Lambda $-term allows to talk about the selforganization of vacuum $%
{}^{22}$. The processes of the creation particles from vacuum energy have
occurred during relativistic phase transitions. The series of RPT must be
continued by the phase transition in dark matter medium. Certainly our
consideration of $\Lambda $-term problem is far from fullness. Zeldovich $%
{}^{21}$ has obtained the formula (2) for $\Lambda $-term using Eddington$%
{}^{23}$ and Dirac${}^{24}$ formulas of large numbers. But this was an
important result at the qualitative level. Nevertheless some conclusions can
be made. There had been phase transitions in the Universe. In the present
epoch the vacuum energy density is a nonzero and positive (vacuum is in a
excited state). The problem of $\Lambda $-term was not solved since which
has been the initial $\Lambda $- term and how it has transformed during
evolution of the Universe we do not know exactly. Probably $\Lambda $-term
is the key to physics of XXI century.

Finally we note, that all the values of fundamental constants, the limits on 
$H_o$ and masses of $\pi $-mesons were taken from the review of particle
properties ${}^{25}$. By tradition the term ''$\Lambda $-term'' is used more
often than the term ''the cosmological constant''. The surprising thing is
that in the article $^{26}$ the square of the most likely values $H_0$ ,$%
\Omega _{\Lambda}$and $\Omega _{m}$ is practically
coincided with our values.

I am grateful to D.Kirzhnits, V.Rubakov, A.Starobinsky and G.Vereshkov for
useful discussion and N.Kardashev and G.Vereshkov for representation of the
manuscripts before the publication.

\vspace*{1cm}

\newpage

\begin{center}
{\bf REFERENCES}\\
\end{center}

\noindent
1. Hawking S.W. The cosmological constant is probably zero. Phys. Lett. B
134, 403-404 (1984).\\ 2. Weinberg S. The cosmological constant problem.
Rev. Mod. Phys. 61, 1-23 (1989).\\ 3. Coleman S. Why there is nothing rather
than something: A theory of the cosmological constant. Nucl. Phys. B 310,
643-658 (1988).\\ 4. Gasperini M. The cosmological constant and the
dimensionality of spacetime. Phys. Lett. B224, 49-52 (1989).\\ 5.
Lavrelashvili G., Rubakov V.A., Tinyakov P.G. Third quantization of gravity
and the cosmological constant problem. Proceedings of the Fifth Seminar
''Quantum Gravity''. Eds. M.A.Markov, V.A.Beresin, V.P.Frolov. World
Scientific, p. 27-43 (1991).\\ 6. Fukugita M. and Yanagida T. How do we
understand a small cosmological constant. Proceedings of the 1 RESCEU
Symposium ''The cosmological constant and the Evolution of the Universe''.
Eds. K.Sato, T.Suginohara, N.Sugiyama. Universal Academy Press, p. 127-131
(1996).\\ 7. Efstathiou G. An anthropic argument for a cosmological
constant. Proceedings of the 1 RESCEU Symposium ''The cosmological constant
and the Evolution of the Universe''. Eds. K.Sato, T.Suginohara, N.Sugiyama.
Universal Academy Press, p. 225-232 (1996).\\ 8. Kolb E.W., Copeland E.,
Liddle A.L., Lidsey J.E., Barreiro T. and Abney M. Deducing the Value of the
Cosmological Constant during Inflation from Present-day
Observations.Proceedings of the 1 RESCEU Symposium ''The cosmological
constant and the Evolution of the Universe''. Eds. K.Sato, T.Suginohara,
N.Sugiyama. Universal Academy Press, p. 169-188 (1996).\\ 9. Vilenkin A.
Quantum Cosmology and the Constants of Nature. Proceedings of the 1 RESCUE
Symposium ''The cosmological constant and the Evolution of the Universe'',
Eds. K.Sato, T.Suginohara, N.Sugiyama. Universal Academy Press, p. 161-168
(1996).\\ 10. Martel H., Shapiro P.R., Weinberg S. Likely value of the
cosmological constant. Astro-ph/9701099.\\ 11. Fukujita M. and Turner E.L.
Gravitational lensing frequencies: galaxy cross sections and selection
effects. MNRAS 253, 99-106 (1991).\\ 12. Krauss L.M. Old galaxies at high
redshifts and the cosmological conststant. Astrophys. J. 480, 466-469 (1997).%
\\ 13. Grishchuk L.P., Zeldovich Ya.B. Complete cosmological theories.
Proceedings of the 2nd Seminar on Quantum Gravity. Eds. M.A.Markov,
P.C.West, Plenum, New York, p.71-86 (1982).\\ 14. Vilenkin A. Creation of
Universes from nothing. Phys. Lett. 117B, 25-28 (1982).\\ 15. Hartle J.B.,
Hawking S.W. Wave function of the Universe. Phys. Rev. D28, 2960- 2975
(1983).\\ 16. Linde A. Quantum creation of inflationary Universe. Lett.
Nuovo Cim. 39, 401-405 (1984).\\ 17. Kardashev N.S. The inflation of the
present Universe. Astron. Zh. (Russia) (submitted) (1997).\\ 18. Burdyuzha
V., Lalakulich O., Ponomarev Yu., Vereshkov G. New scenario for the early
evolution of the Universe. Phys. Rev.D55,7340R-7345R (1997).\\ 19. Linde
A.D. Particle Physics and Inflationary Cosmology. ,Eds. Harwood, Chur,
Switzerland (1990).\\ 20. Kolb E.W. Cosmological phase transitions.
Proceedings Yamada Conference XXXVII ''Evolution of the Universe and its
Observational Quest''. Ed. K.Sato. Yamada Science Foundation and Universal
Academy Press, p. 31-48 (1994).\\ 21. Zeldovich Ya.B. The cosmological
constant and elementary particles. Pis'ma JETP 6, 883-884 (1967).\\ 22.
Vereshkov G.M. The vacuum physics and $\Lambda $-term problem. Preprint of
Lebedev Physical Inst. of Russian Academy of Sciences (submitted) (1997).\\ %
23. Eddington F.R. On the value of the cosmical constant. Proc. Roy. Soc.
London 133, 605-615 (1931).\\ 24. Dirac P.A.M/ A new basis for cosmology.
Proc. Roy. Soc. London 165, 199-208 (1938).\\ 25. Barnett R.M. et al. Review
of Particle Properties. Phys. Rev. D54, 65 and 112 (1996).\\  26. Ostriker
J.P., Steinhardt P.J. The observational case for a low density Universe with
a non-zero cosmological constant. Nature 377, 600-603 (1995). 

\end{document}